\documentclass[journal,transmag]{IEEEtran}

\usepackage{hyperref}

\ifCLASSINFOpdf
   \usepackage[pdftex]{graphicx}
\else
\fi
\usepackage{amsmath}
\usepackage{amsfonts}
\ifCLASSOPTIONcompsoc
  \usepackage[caption=false,font=normalsize,labelfont=sf,textfont=sf]{subfig}
\else
  \usepackage[caption=false,font=footnotesize]{subfig}
\fi

\usepackage[export]{adjustbox}
\usepackage{siunitx}
\usepackage{enumitem}
\usepackage{soul}
\usepackage[straightvoltages]{circuitikz}
\usepackage{tikz}
\usetikzlibrary{decorations.pathreplacing}
\usepackage[backend=biber,sorting=none,style=ieee]{biblatex}

\begin{document}
\title{ Complex Reluctivity and Dynamic Hysteresis Model 
for Soft Magnetic Composite Materials}


\author{\IEEEauthorblockN{Joonas Vesa\IEEEauthorrefmark{}}
%Hajime Igarashi\IEEEauthorrefmark{2} }
\IEEEauthorblockA{\IEEEauthorrefmark{}Electrical Engineering Unit,
Tampere University, Tampere 33014, Finland}
%\IEEEauthorblockA{\IEEEauthorrefmark{2} Graduate School of Information 
%Science and Technology, Hokkaido University, Sapporo 061-0814, Japan  }

%\thanks{Manuscript received Dec. 18th, 2023; revised *** *th, 202*. 
%Corresponding author: J. Vesa 
%(\url{https://researchportal.tuni.fi/fi/persons/joonas-vesa}).}
}

%\author{\IEEEauthorblockN{Iiiiiiii Iiiiiiii\IEEEauthorrefmark{1} and
%Iiiiiiii Iiiiiiii\IEEEauthorrefmark{2} }
%\IEEEauthorblockA{\IEEEauthorrefmark{1,2}Iiiiiiiiiii Ii Iiiiiiii Iiiiiiiii Iiii, 
%Iiiii Iiiiiiiiii ii Iiiiiiiiii, Iiii, IIIIIII Iiiiiiiii}
%%\IEEEauthorblockA{\IEEEauthorrefmark{2}Twentieth Century Fox, Springfield, USA}
%
%\thanks{Manuscript received December 1, 2012; revised August 26, 2015. 
%Corresponding author: I. Iiiiiiii (email: google.com).}}

\markboth{}%IEEE,~Vol.~?, No.~?, ??????~????}%
{Vesa \MakeLowercase{\textit{et al.}}: }
%{Iiiiiiii \MakeLowercase{\textit{et al.}}: }



%123456789 123456789 123456789 123456789 123456789 123456789 123456789 123456789


% for Transactions on Magnetics papers, we must declare the abstract and
% index terms PRIOR to the title within the \IEEEtitleabstractindextext
% IEEEtran command as these need to go into the title area created by
\IEEEtitleabstractindextext{%
\begin{abstract}
%\hli{In this paper, for the first time in the literature,
%analytical closed-form coefficients 
%are derived for a Cauer ladder reluctivity model of 
%soft magnetic composites. The derived coefficients depend
%on the physical and geometric characteristics
%of the material, and no experimental fitting parameters are used. 
%}
%
In this article, we derive a Cauer network reluctivity model
for soft magnetic composite materials. 
The Cauer network components are closed-form expressions consisting
only of such parameters that have a clear physical meaning.
The parameters include particle permeability, particle
conductivity, insulation permeability, insulation thickness, 
volume fraction,
mean particle diameter, variance of particle diameters, and maximal
particle diameter.
The particle volumes are assumed to be beta-distributed.
Since the model is a network, it is immediately available in both
frequency- and time domains.
\end{abstract}

%\begin{IEEEkeywords}
%Cauer network, complex reluctivity, continued fraction, 
%dynamic hysteresis, effective radius, eddy current, 
%particle volume distribution, soft magnetic composite.
%\end{IEEEkeywords}
}

\maketitle

\IEEEdisplaynontitleabstractindextext

% \fi
\IEEEpeerreviewmaketitle

\section{Introduction}
\label{Section-Preliminaries}
Soft magnetic composite (SMC) materials consist of electrically
insulated and compacted ferromagnetic particles \cite{shokrollahi_soft_2007}.
SMC materials provide isotropy
and reduction of eddy currents. Relative permeabilities of such materials 
tend to be low, of the order of a hundred.

As a starting point for our analysis, we use the complex 
reluctivity model of SMC materials used in the literature 
\cite{vesa_eddy-current_2023}.
The reluctivity reads as
\begin{equation}
\label{nu_ref}
\begin{cases}
\nu_l(\omega) &= \frac{1}{(1+\tau)\eta}\left [ 
\frac{\tau}{\mu_\text{ins}} 
+ \frac{1}{\mu} \frac{R(l)k(\omega)J_0(R(l)k(\omega))}{2J_1(R(l)k(\omega))}
\right ], \\
\nu_\text{ref}(\omega) &= 
\frac{\int_L \nu_l(\omega) l g(l) \text{d}l }
		   {\int_L l g(l) \text{d}l },
\end{cases}
\end{equation}
where 
\begin{itemize}
\item
$\nu_l$ is the reluctivity of a single particle,
\item
$l$ is the length of a single particle in the 
direction of the magnetic field,
\item
$\omega$ is angular frequency $\omega = 2\pi f$, 
where $f$ is frequency,
\item
$\tau$ denotes the relative insulation thickness of 
the particles, which is the ratio between the 
insulation thickness and the particle length,
\item
$\eta$ is the volume fraction of the material,
\item
$\mu_\text{ins}$ is the permeability of the insulations,
\item
$\mu$ is the permeability of the particles,
\item
$R(l)=l/(2\kappa)$ is the particle radius of a single particle
in the direction perpendicular
to the magnetic field, where
\item
$\kappa$ is the ratio of the length $l$ and the diameter $2R$ 
of the particles.
\item
$k(\omega)$ is a short-hand notation for the expression 
$\sqrt{-\mu\sigma j\omega}$, where
\item
$\sigma$ is the conductivity of the particles, and
\item
$j$ is the imaginary unit.
\item
$J_0$ is the zeroth order Bessel function of the first kind,
\item
$J_1$ is the first order Bessel function of the first kind,
\item
$\nu_\text{ref}$ is the reluctivity of the whole material,
\item
$g$ is a distribution of particle sizes expressing 
"how many particles have the length $l$",
\item
$L$ is the domain of the particle lengths of the distribution $g$.
\end{itemize}

\section{Objectives}
We want to express \eqref{nu_ref} as a continued
fraction, interpreted as a Cauer network, 
with each component having a closed-form analytical expression.
This is a cumbersome task, since computing the integrals analytically is not
straightforward. Hence, we approximate \eqref{nu_ref} by
considering 
\begin{equation}
\label{nu_eff}
\nu_\text{eff}(\omega) = \frac{1}{(1+\tau)\eta}\left [ 
\frac{\tau}{\mu_\text{ins}} 
+ \frac{1}{\mu} 
\frac{R_\text{eff}k(\omega)J_0(R_\text{eff} k(\omega))}
{2J_1(R_\text{eff} k(\omega))}
\right ],
\end{equation}
and solving the effective radius $R_\text{eff}$ under the conditions
\begin{eqnarray}
\label{ehto1}
\lim_{\omega \rightarrow 0^+} \nu_\text{ref}(\omega)
&=& \lim_{\omega \rightarrow 0^+} 
\nu_\text{eff}(\omega) \in \mathbb{C}^{\neq 0}, \\
\label{ehto2}
\lim_{\omega \rightarrow \infty} \frac{1}{k(\omega)}\nu_\text{ref}(\omega)
&=& \lim_{\omega \rightarrow \infty} \frac{1}{k(\omega)}\nu_\text{eff}(\omega)
\in \mathbb{C}^{\neq 0}.
\end{eqnarray}

We now adopt the following terminology; the model \eqref{nu_ref} is called
\emph{reference} model and \eqref{nu_eff} is called \emph{effective} model.
Conditions \eqref{ehto1} and \eqref{ehto2} ensure that
the reference reluctivity and the effective reluctivity agree 
at the DC-limit, as well as asymptotically.
We require that both limits yield non-zero non-infinite complex numbers. 
We have farsightedly included the coefficient $1/k(\omega)$ 
in the second condition to divide off the rate of increase from 
the reluctivities. The interpretation is that we seek for such a particle 
radius $R_\text{eff}$ that allows us to model the whole material using the 
single-particle model \eqref{nu_eff}. We will see that $R_\text{eff}$ 
can be uniquely determined by these conditions.

After finding $R_\text{eff}$, we expand the effective model $\eqref{nu_eff}$
as a Gauss continued fraction. The continued fraction corresponds to a
Cauer circuit, which approximates the dynamics of \eqref{nu_ref}
in the sense of \eqref{ehto1} and \eqref{ehto2}.

\section{Effective Radius}
\label{Section-EffectiveRadius}
In this section, we find an expression for the effective radius 
$R_\text{eff}$ and we show that conditions \eqref{ehto1} and \eqref{ehto2}
hold. We need to define the particle volume density first.

\subsection{Particle Volume Density}
We make a terminological distinction between particle \emph{length} in
the direction of the magnetic field and particle \emph{diameter} in the 
direction perpendicular to the magnetic field. 
Of course, usually these quantities are equal
and we set $\kappa=1$.

Roughly speaking, a particle volume density expresses
"what percentage of the total volume is occupied by
particles of length $l$".
We use the beta distribution to model particle volume 
densities with respect to particle lengths. The beta 
distribution has the form
\begin{equation}
\label{betadistr}
G_{\alpha,\beta,l_\text{max}}(l) = Dl^{\alpha-1}(l_\text{max}-l)^{\beta-1},
\end{equation}
where $\alpha$ and $\beta$ are shape parameters of the density function and
$l_\text{max}$ is a maximal particle length 
\cite{chattamvelli_continuous_2021}.
$D$ is a normalization coefficient. The support of $G$ is 
$[0,l_\text{max}]$, and $G$ is defined for 
\begin{equation}
\label{conditions1}
\begin{cases}
\alpha > 0, \\
\beta > 0, \\
l_\text{max} > 0,
\end{cases}
\end{equation}
to ensure integrability of the density function.
Later, we give more restrictions for $\alpha$, $\beta$, and $l_\text{max}$.

The shape parameters $\alpha$ and $\beta$ are related 
to the expected particle length $\mu_G$, 
the variance of particle lengths $\sigma^2_G$, and
the maximal particle length $l_\text{max}$ of the 
particle volume distribution \eqref{betadistr} by the formulas
\begin{equation}
\label{alpha}
\alpha = \frac{(l_\text{max}\mu_G-\mu_G^2-\sigma_G^2)\mu_G }
		{l_\text{max}\sigma_G^2 }
\end{equation}
and
\begin{equation}
\label{beta}
\beta = \frac{
	(l_\text{max}\mu_G-\mu_G^2-\sigma_G^2) (l_\text{max}-\mu_G)
		}
		{l_\text{max}\sigma_G^2 }. 
\end{equation}
The subscript $G$ in the symbols $\mu_G$ and $\sigma_G^2$ prevent us
from confusing these statistical parameters with permeability and conductivity.

For concrete computations, particle numbers are needed
in \eqref{nu_ref}. 
Since particle volumes are proportional to $l^3$, a density function 
for particle numbers is given by 
\begin{equation}
\label{numdistr}
g_{\alpha,\beta,l_\text{max}}(l) = Cl^{\alpha-4}(l_\text{max}-l)^{\beta-1},
\end{equation}
where $C$ is a normalization coefficient. 
Roughly speaking, the density function $g$ expresses "how many particles 
have the length $l$".
The shape parameters $\alpha$ and $\beta$ are given in \eqref{alpha}
and \eqref{beta}, but we must remember that the corresponding statistical 
parameters $\mu_G$ and $\sigma_G^2$ refer to the particle volume density $G$
and not the particle number density $g$.
We will use \eqref{numdistr} in \eqref{nu_ref} with the shape parameters
\eqref{alpha} and \eqref{beta}.

Next, we check that the conditions \eqref{ehto1} 
and \eqref{ehto2} hold, and we define the effective
radius $R_\text{eff}$ for the effective model \eqref{nu_eff} accordingly. 
The next two subsections contain rather technical computations. 
We summarize the results briefly here, and suggest 
application-oriented readers to skip Sections \ref{lfsection} 
and \ref{hfsection}.
In Section \ref{lfsection}, we find that the condition \eqref{ehto1} holds, 
provided that constraints \eqref{conditions3} hold.
In Section \ref{hfsection}, we find that the condition \eqref{ehto2} holds 
if and only if $R_\text{eff}$ is defined as in \eqref{reffdef} and if the 
constraints \eqref{conditions4} hold.

\subsection{Low-frequency Limit}
\label{lfsection}
We check that the condition \eqref{ehto1} holds for 
\eqref{nu_ref} and \eqref{nu_eff}. 
It means that the reluctances of both models
agree at the DC-limit.

We compute the limit for $\nu_\text{eff}$ first. On a neighbourhood around 
the origin of the complex plane, we have the power series expansion
\begin{equation}
\label{powerseries}
\frac{zJ_0(z)}{2J_1(z)} = 1 + \sum_{i=1}^\infty \gamma_{2i}z^{2i},
\end{equation}
where $2i = -1/8, -1/192, ...$. Hence, we have
\begin{equation}
\begin{split}
& \lim_{\omega \rightarrow 0^+} 
\frac{Rk(\omega)J_0(Rk(\omega))}{2J_1(Rk(\omega))} \\
& \ \ \ \ \ \ = 1 + \lim_{\omega \rightarrow 0^+} 
\sum_{i=1}^\infty \gamma_{2i}(Rk(\omega))^{2i} = 1,
\end{split}
\end{equation}
and thus,
\begin{equation}
\label{nuefflim0}
\begin{split}
\lim_{\omega \rightarrow 0^+} \nu_\text{eff}(\omega) &=
\frac{1}{(1+\tau)\eta}\bigg [ 
\frac{\tau}{\mu_\text{ins}} \\
& \ \ \ \ \ \ + \frac{1}{\mu} \lim_{\omega \rightarrow 0^+}
\frac{R_\text{eff}k(\omega)J_0(R_\text{eff} k(\omega))}
{2J_1(R_\text{eff} k(\omega))}
\bigg ] \\
%&=\frac{1}{(1+\tau)\eta}\bigg [ 
%\frac{\tau}{\mu_\text{ins}} + \frac{1}{\mu} \\
%& \ \ \ \ \ \ + \frac{1}{\mu}\lim_{\omega \rightarrow 0^+}
%\sum_{i=1}^\infty \gamma_{2i}(Rk(\omega))^{2i}
%\bigg ] \\
&= \frac{1}{(1+\tau)\eta}\left [ 
\frac{\tau}{\mu_\text{ins}} 
+ \frac{1}{\mu} \right ]. \\
\end{split}
\end{equation}

Next, we compute the limit for $\nu_\text{ref}$.
We get
\begin{equation}
\begin{split}
&\lim_{\omega \rightarrow 0^+} \nu_\text{ref}(\omega) \\
& = \lim_{\omega \rightarrow 0^+} 
	\frac{\int_L \nu_l(\omega) l g_{\alpha,\beta,l_\text{max}}(l) \text{d}l }
	   {\int_L l g_{\alpha,\beta,l_\text{max}}(l) \text{d}l } \\
&=\frac{1}{(1+\tau)\eta}\bigg [ 
\frac{\tau}{\mu_\text{ins}} \\
& \ \  + \frac{1}{\mu} 
\lim_{\omega \rightarrow 0^+} 
\frac{\int_L \frac{R(l)k(\omega)J_0(R(l)k(\omega))}{2J_1(R(l)k(\omega))} 
l g_{\alpha,\beta,l_\text{max}}(l) \text{d}l }
	   {\int_L l g_{\alpha,\beta,l_\text{max}}(l) \text{d}l }
\bigg ] \\
&=
\frac{1}{(1+\tau)\eta}\bigg [ 
\frac{\tau}{\mu_\text{ins}} 
+ \frac{1}{\mu}  \\
& \ \  + \frac{1}{\mu}\lim_{\omega \rightarrow 0^+} \frac{
\int_L \sum_{i=1}^\infty \gamma_{2i}(R(l)k(\omega))^{2i} 
l g_{\alpha,\beta,l_\text{max}}(l) \text{d}l }
		   {\int_L l g_{\alpha,\beta,l_\text{max}}(l) \text{d}l }
\bigg ].
\end{split}
\end{equation}
In a small enough neighbourhood of the origin, 
the integrand in the numerator is bounded by the function $lg(l)$. 
In Appendix B we integrate such functions, and we find that
the bound is integrable if
\begin{equation}
\label{conditions3}
\begin{cases}
\alpha > 2, \\
\beta > 0, \\
l_\text{max} > 0.
\end{cases}
\end{equation}
Since an integrable bound is found, the limit can be taken under the 
integral sign  by the dominated convergence theorem, and the integral 
term vanishes \cite{rudin_real_1986}. We get
\begin{equation}
\lim_{\omega \rightarrow 0^+} \nu_\text{ref}(\omega) = 
\frac{1}{(1+\tau)\eta}\left [ 
\frac{\tau}{\mu_\text{ins}} 
+ \frac{1}{\mu} \right ],
\end{equation}
which is equal to \eqref{nuefflim0} and concludes the consideration. 
The condition \eqref{ehto1} holds,
but we had to make an additional constraint \eqref{conditions3} 
for $\alpha$ compared to what was given in \ref{conditions1}.

\subsection{High-frequency Limit}
\label{hfsection}
We check that the condition \eqref{ehto2} holds for \eqref{nu_ref} and 
\eqref{nu_eff}, provided that particle sizes are modelled using 
\eqref{betadistr} (and thus, \eqref{numdistr}).
We check that the reluctances agree asymptotically. 
We define $R_\text{eff}$ accordingly.

Let us first consider the asymptotic limit of $\nu_\text{eff}$.
In Appendix A, we show that
\begin{equation}
\label{besselinf}
\lim_{\omega \rightarrow \infty}
\frac{J_0(R(l)k(\omega))}{J_1(R(l)k(\omega))} = j.
\end{equation}
In the following, we will use the fact that 
$\lim_{\omega\rightarrow\infty}1/k(\omega)=0$,
and we will substitute \eqref{besselinf}. 
By a straightforward computation, the asymptotic limit of 
\eqref{nu_eff} turns into
\begin{equation}
\label{nueffinf}
\begin{split}
\lim_{\omega\rightarrow\infty} \frac{1}{k(\omega)}\nu_\text{eff}
&= \lim_{\omega\rightarrow\infty} \frac{1}{k(\omega)}
\frac{1}{(1+\tau)\eta}\bigg [ 
\frac{\tau}{\mu_\text{ins}} \\
& \ \ \ +\frac{1}{\mu} 
\frac{R_\text{eff}k(\omega) J_0(R_\text{eff}k(\omega))}
{2J_1(R_\text{eff}k(\omega))} \bigg ] \\
&= \frac{1}{(1+\tau)\eta} 
\frac{1}{\mu} 
\frac{R_\text{eff}}{2} 
\lim_{\omega\rightarrow\infty}
\frac{J_0(R_\text{eff}k(\omega))}{J_1(R_\text{eff}k(\omega))} \\
&= \frac{j}{2\mu(1+\tau)\eta}R_\text{eff}.
\end{split}
\end{equation}

More complications arise when considering the asymptotic limit of 
$\nu_\text{ref}$, since the calculations involve taking a limit under 
an integral sign. First of all, the asymptotic limit of
\eqref{nu_ref} turns into
\begin{equation}
\label{highfreq1}
\begin{split}
&
\lim_{\omega\rightarrow\infty} \frac{1}{k(\omega)}\nu_\text{ref} \\
&= \lim_{\omega\rightarrow\infty} \frac{1}{k(\omega)}
\frac{1}{(1+\tau)\eta}\bigg [ 
\frac{\tau}{\mu_\text{ins}} \\
& \ \ \ \ \ +\frac{1}{\mu} 
\frac{\int_L
\frac{R(l)k(\omega) J_0(R(l)k(\omega))}
{2J_1(R(l)k(\omega))}l g_{\alpha,\beta,l_\text{max}}(l) \text{d}l
}{\int_L
l g_{\alpha,\beta,l_\text{max}}(l) \text{d}l} \bigg ] \\
&= 
\frac{1}{(1+\tau)\eta}\bigg [\frac{1}{\mu} 
\lim_{\omega\rightarrow\infty} \frac{1}{k(\omega)} \\
& \ \ \ \ \ \cdot \frac{\int_L
\frac{R(l)k(\omega) J_0(R(l)k(\omega))}
{2J_1(R(l)k(\omega))}l g_{\alpha,\beta,l_\text{max}}(l) \text{d}l
}{\int_L
l g_{\alpha,\beta,l_\text{max}}(l) \text{d}l} \bigg ]. \\
\end{split}
\end{equation}
The limit only acts on the numerator, since the
denominator has no $\omega$-dependency. We compute 
\begin{equation}
\label{highfreq2}
\begin{split}
&\lim_{\omega\rightarrow\infty} \frac{1}{k(\omega)}
\int_L
\frac{R(l)k(\omega) J_0(R(l)k(\omega))}
{2J_1(R(l)k(\omega))}l g_{\alpha,\beta,l_\text{max}}(l) \text{d}l \\
&= \lim_{\omega\rightarrow\infty} 
\int_L
\underbrace{
\frac{R(l)J_0(R(l)k(\omega))}
{2J_1(R(l)k(\omega))}}_{=:Y(l,\omega)}  l g_{\alpha,\beta,l_\text{max}}(l) 
\text{d}l. \\
\end{split}
\end{equation}
In order to take the limit under the integral sign, we will
argue that the function $Y(l,\omega)l g_{\alpha,\beta,l_\text{max}}(l)$ 
is bounded by an integrable function, when $\omega_0<\omega$ for some 
positive $\omega_0$. 
Then, we can use the dominated convergence theorem
\cite{rudin_real_1986}
to switch the order of the limit and the integration.

Let us first argue that the function $Y$ can be reparameterized
into a continuous function on a compact domain. 
\begin{itemize}
\item
When considering the limit, the domain of $Y$ can be set
to $(0,l_\text{max}]\times[\omega_0,\infty)$,
where the positive $w_0$ can be freely chosen.
\item
For $\omega_0<\omega$, using \eqref{powerseries}, we have 
\begin{equation}
\begin{split}
\lim_{l\rightarrow 0^+}Y(l,w) 
&= \lim_{l\rightarrow 0^+}
\frac{R(l)J_0(R(l)k(\omega))}{2J_1(R(l)k(\omega))} \\
&= \lim_{l\rightarrow 0^+} \frac{1}{k(\omega)}
\frac{R(l)k(\omega)J_0(R(l)k(\omega))}{2J_1(R(l)k(\omega))} \\ 
&= \frac{1}{k(\omega)} \lim_{l\rightarrow 0^+}
\left ( 1 + \sum_{i=1}^\infty (R(l)k(\omega))^{2i} \right ) \\ 
&= \frac{1}{k(\omega)},
\end{split}
\end{equation}
which is continuous in $\omega$. 
Hence, $Y$ can be continuously extended
into $[0,l_\text{max}]\times[\omega_0,\infty)$.
\item
Using \eqref{besselinf}, we have the limit
\begin{equation}
\begin{split}
\lim_{\omega\rightarrow \infty}Y(l,w) 
&= \lim_{\omega\rightarrow \infty}
\frac{R(l)J_0(R(l)k(\omega))}{2J_1(R(l)k(\omega))} \\
&=\frac{R(l)j}{2}, 
\end{split}
\end{equation}
which is continuous in $l$. 
We can reparameterize the continuous extension
of $Y$ into a compact domain 
$[0,l_\text{max}]\times[\omega_0,\omega_1]$, 
where $\omega_0<\omega_1$.
As a continuous function, it must be bounded. 
\end{itemize}
We have concluded that $Y$ is bounded.
If we denote the bound of $Y$ by $M$, the function 
$Y(l,\omega)l g_{\alpha,\beta,l_\text{max}}(l)$
is bounded by $Ml g_{\alpha,\beta,l_\text{max}}(l)$,
which is integrable under the conditions \eqref{conditions3}, 
as indicated in Appendix B. 
Hence, we may take the limit 
\eqref{highfreq2} under the integral 
by the dominated convergence theorem \cite{rudin_real_1986}.
Substituting \eqref{besselinf}, we get
\begin{equation}
\label{highfreq3}
\begin{split}
&\lim_{\omega\rightarrow\infty}
\int_L
\frac{R(l) J_0(R(l)k(\omega))}
{2J_1(R(l)k(\omega))}l g_{\alpha,\beta,l_\text{max}}(l) \text{d}l \\
&= 
\int_L
\lim_{\omega\rightarrow\infty} 
\frac{R(l)J_0(R(l)k(\omega))}
{2J_1(R(l)k(\omega))}  l g_{\alpha,\beta,l_\text{max}}(l) \text{d}l \\
&= 
\frac{j}{2} \int_L
R(l) l g_{\alpha,\beta,l_\text{max}}(l) \text{d}l. \\
\end{split}
\end{equation}
Collecting the results of \eqref{highfreq1} and \eqref{highfreq3}, we have
\begin{equation}
\label{highfreq4}
\begin{split}
&
\lim_{\omega\rightarrow\infty} \frac{1}{k(\omega)}\nu_\text{ref} 
= \frac{1}{(1+\tau)\eta} \frac{1}{\mu} 
\frac{
\frac{j}{2} \int_L
R(l) l g_{\alpha,\beta,l_\text{max}}(l) \text{d}l}{\int_L
l g_{\alpha,\beta,l_\text{max}}(l) \text{d}l} . \\
\end{split}
\end{equation}
In Appendix B, we show that
\begin{equation}
\frac{\int_L R(l) l g_{\alpha,\beta,l_\text{max}}(l) \text{d}l }
		   {\int_L l g_{\alpha,\beta,l_\text{max}}(l) \text{d}l } = 
		   \frac{l_\text{max}}{2\kappa}\frac{\alpha-2}{\alpha+\beta-2}
\end{equation}
under the conditions \eqref{conditions3}. 
Computing the integrals of 
\eqref{highfreq4} with this integral formula, we finally have
\begin{equation}
\label{highfreq5}
\begin{split}
&
\lim_{\omega\rightarrow\infty} \frac{1}{k(\omega)}\nu_\text{ref} 
= \frac{j}{2\mu(1+\tau)\eta}  
\frac{l_\text{max}}{2\kappa}\frac{\alpha-2}{\alpha+\beta-2}. \\
\end{split}
\end{equation}
Comparing with \eqref{nueffinf}, the asymptotic agreement \eqref{ehto2} 
between $\nu_\text{eff}$ and $\nu_\text{ref}$ holds if and only if we define
\begin{equation}
\label{reffdef}
R_\text{eff} = \frac{l_\text{max}}{2\kappa}\frac{\alpha-2}{\alpha+\beta-2}.
\end{equation}
The parameters $l_\text{max}$, $\alpha$, and $\beta$ are constrained by
\begin{equation}
\label{conditions4}
\begin{cases}
\alpha > 2, \\
\beta > 0, \\
l_\text{max} > 0,
\end{cases}
\end{equation}
since we referred to similar conditions 
\eqref{conditions3} in this section. 
Other combinations of the parameters may still 
provide a positive radius,
but if conditions \eqref{conditions4} are not met, we do not
guarantee the properties \eqref{ehto1} and 
\eqref{ehto2} to hold.

\subsection{Further Manipulations}
The effective radius \eqref{reffdef} depends on the 
maximal particle length 
$l_\text{max}$ and the shape parameters $\alpha$ and 
$\beta$ of the particle 
volume density \eqref{betadistr}. 
Even though some readers may find these parameters intuitive, 
we want to express the radius in terms
of the maximal particle length $l_\text{max}$, 
the expected particle length $\mu_G$, 
and the variance of particle lengths $\sigma_G^2$ of 
a beta-distributed volume density.

Substituting \eqref{alpha} and \eqref{beta} 
into \eqref{reffdef} 
and performing obvious manipulations yields
\begin{equation}
\label{reffdef2}
R_\text{eff} = \frac{1}{2\kappa} 
\frac{(l_\text{max}-\mu_G)\mu_G^2-(2l_\text{max}+\mu_G)\sigma_G^2}
{(l_\text{max}-\mu_G)\mu_G-3\sigma_G^2}.
\end{equation}
Constraints that guarantee \eqref{ehto1} and \eqref{ehto2} 
are given by
\begin{equation}
\label{conditions5}
\begin{cases}
(\sigma_G^2 > 0), \\
\mu_G^2 > 2\sigma_G^2, \\
l_\text{max} > \frac{\mu_G^2+\sigma_G^2}{\mu_G-\frac{2\sigma_G^2}{\mu_G}},
\end{cases}
\end{equation}
that are equivalent to the constraints \eqref{conditions4}. 
The equivalency of these constraints is shown in Appendix C. 
If one wants to use the effective radius \eqref{reffdef2} outside the 
conditions \eqref{conditions5}, 
the radius $R_\text{eff}$ may still be positive,
but in that case, we do not guarantee the properties 
\eqref{ehto1} and \eqref{ehto2} to hold.
We put the first condition inside parentheses, 
since taking the limit 
$\sigma_G^2\rightarrow 0$ yields
\begin{equation}
\label{reffdef3}
R_\text{eff} \rightarrow \frac{\mu_G}{2\kappa},
\end{equation}
which is physically understandable. 
In this case, the effective length of the particles $2\kappa R_\text{eff}$ 
coincides with the expected particle length. 
Hence, we allow $\sigma_G^2=0$ as well, even though we did not
provide a mathematical proof for the models \eqref{nu_ref} and \eqref{nu_eff} 
to be equal under the limit $\sigma_G^2 \rightarrow 0$. 
We let the physical argument suffice now.

If $l_\text{max}$ is large, or if \eqref{reffdef2} feels too complicated, 
one may want to take a limit $l_\text{max} \rightarrow \infty$. 
Dividing both the numerator and the denominator of \eqref{reffdef2} 
by $l_\text{max}-\mu_G$ and computing the limits yields
\begin{equation}
\label{reffdef4}
R_\text{eff} \rightarrow \frac{1}{2\kappa} 
\left ( \mu_G - 2 \frac{\sigma_G^2}{\mu_G} \right ),
\end{equation}
with the conditions 
\begin{equation}
\label{conditions6}
\begin{cases}
\mu_G > 0, \\
\mu_G^2 > 2\sigma_G^2 \\
\end{cases}
\end{equation}
keeping the expected particle length and the radius positive.

\section{Continued Fration and Cauer Network}
At this point, the model \eqref{nu_eff} together with the 
effective radius \eqref{reffdef2} and the constraints \eqref{conditions5}
(or radius \eqref{reffdef4} and constraints \eqref{conditions6})
provide a magnetic model in the frequency domain. 
For time domain simulations, additional trickery is needed. 
We turn the frequency domain model into
a continued fraction and interpret it as a
Cauer network, which is available in both
frequency- and time domains.

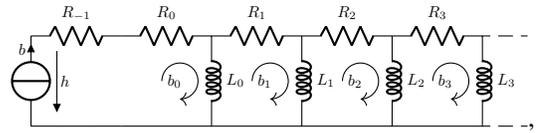
\begin{figure}[h]
	\centering
  \begin{circuitikz}[scale=0.6, transform shape ]
  \draw
    (0,-2) to[I=$b$, v=$h$] (0,0)
    (0,0) to[R=$R_{-1}$] (2,0) coordinate(tmp1)
    (0,-2) to[short] (2,-2) coordinate(tmp2);
    \foreach\n in {0,1,2,3} { 
	\path (tmp1) to ++(0.9,-1) coordinate(tmp0);
	\draw[->] (tmp0) arc(180:-90:0.4) node[pos=0,right] {$b_{\n}$};
	\draw
	(tmp2) to[short] ++(2,0) coordinate(tmp2)
  (tmp1) to[R=$R_{\n}$] ++(2,0) coordinate(tmp1)
  (tmp1) to[L=$L_{\n}$] ++(0,-2) coordinate(tmp2);
    }  
    \draw[dashed] (tmp1) to[short] ++(1,0)
    (tmp2) to[short] ++(1,0)
; 
\end{circuitikz},
\caption{Cauer network representation of \eqref{nu_eff_cf}.}
 \label{cauerladder}
\end{figure}

Using the recurrence formula 
\cite{abramowitz_handbook_1972}
\begin{equation}
  \frac{2\nu}{z} J_\nu(z)= J_{\nu-1}(z) + J_{\nu+1}(z),
\end{equation}
we find the expansion \cite{abramowitz_handbook_1972}
\begin{equation}
  \label{ekspansio}
\begin{split}
\frac{J_\nu(z)}{ zJ_{\nu-1}(z) } 
%&= 
%\cfrac{1}{ 2\nu - z^2\cfrac{J_{\nu+1}(z)}{zJ_\nu(z)}} \\
%&= \cfrac{1}{ 2\nu + \cfrac{-z^2}{ 2(\nu+1) - 
%z^2\cfrac{J_{\nu+2}(z)}{zJ_{\nu+1}(z)}}  } \\
&= \cfrac{1}{ 2\nu + \cfrac{-z^2}{ 2(\nu+1) +
\cfrac{-z^2}{ 2(\nu+3) + \cfrac{-z^2}{ \ddots } } }}. \\
\end{split}
\end{equation}
By plugging \eqref{ekspansio} into \eqref{nu_eff}
and performing obvious manipulations, we find
\begin{equation}
\label{nu_eff_cf}
\begin{split}
&\nu_{\text{eff}} 
=
\frac{\tau}{\mu_\text{ins}(1+\tau)\eta} + 
\frac{1}{\mu(1+\tau)\eta} \\
& + \frac{1}{  
\frac{(8\cdot 1)(1+\tau)\eta }{\sigma R_\text{eff}^2 j\omega }
+ \frac{1}{  
\frac{3}{\mu(1+\tau)\eta}
+ \frac{1}{  
\frac{(8\cdot 2)(1+\tau)\eta}{\sigma R_\text{eff}^2 j\omega}
+ \frac{1}{  
\frac{5}{\mu(1+\tau)\eta}
+ {\tiny \cdots}
}
}
}
}.
\end{split}
\end{equation}
This expansion has the interpretation of a Cauer ladder,
depicted in Fig. \ref{cauerladder}.
 \\
The circuit components are given as
\begin{align}
  \label{resistances}
  R_{-1}&=\frac{\tau}{\mu_\text{ins}(1+\tau)\eta}, \\
  R_{i} &=\frac{1+2i}{\mu(1+\tau)\eta}, \\
  \label{inductances}
  L_{i} &=\frac{\sigma R_\text{eff}^2}{8(i+1)(1+\tau)\eta},
\end{align}
with $i = 0,1,2,...$.

\begin{table}[t]
	\caption{Summary of Quantities}
	\begin{center}
	\begin{tabular}{c c}
	Quantity & \text{Description} \\
	\hline\hline
	$\mu_G$ & Particle length mean\\
	$\sigma_G^2$ & Particle length variance\\
	$l_\text{max}$ & Maximal particle length\\
	$\kappa$ & Relative particle length\\
	$\eta$ & Volume fraction (filling factor)\\
	$\mu_\text{ins}$ & Insulation permeability\\
	$\tau$ & Relative insulation thickness \\
	$\mu$ & Particle permeability\\
	$\sigma$ & Particle conductivity\\
	\label{taulukko}
	\end{tabular}
	\end{center}
  \end{table}

\section{Conclusion}
We propose to model SMC materials using the
Cauer ladder, presented in 
Fig. \ref{cauerladder} and 
\eqref{resistances}-\eqref{inductances},
together with the newly derived effective particle radius 
\eqref{reffdef2} under the conditions \eqref{conditions5}.
An easier, but approximated alternative for the radius and the conditions are given 
by \eqref{reffdef4} and \eqref{conditions6}.
Explanations of the symbols can be found 
in Table \ref{taulukko}.
The model is immediately available in time-
and frequency domains.

\section*{Acknowledgment}
This study was mainly funded by the foundation of Ulla Tuominen,
a member of the Foundations' Post Doc Pool. This project has received
funding from the European Research Council (ERC) under the European
Union's Horizon 2020 research and innovation programme (grant agreement
No 848590). The Academy of Finland is also acknowledged for financial
support (grant No 330062).

\section*{Appendix A}
We show that
\begin{equation}
\label{lemma1}
\lim_{\omega \rightarrow \infty}
\frac{J_0(R(l)k(\omega))}{J_1(R(l)k(\omega))} = j
\end{equation}
for any $l>0$.

In the literature \cite{abramowitz_handbook_1972}, 
it is stated for constant angles $-\pi<\arg z<0$, that
\begin{equation}
\label{asymptotic_bessel}
J_\alpha(z) \sim \frac{1}{\sqrt{2\pi z}} 
e^{j(z-\frac{\alpha\pi}{2}-\frac{\pi}{4})}
\bigg ( =: A_\alpha(z) \bigg )
\end{equation}
for Bessel functions $J_\alpha$ of the first kind and order
$\alpha$.
The asymptotic limit $\sim$ 
can be understood in the sense that
\begin{equation}
\label{limJA}
\lim_{|z| \rightarrow \infty} \frac{J_\alpha(z)}{A_\alpha(z)} = 1.
\end{equation}
Before touching the limit \eqref{lemma1}, 
we notice that a straightforward computation gives
\begin{equation}
\begin{split}
\label{A-fraction}
\frac{jA_1(z)}{A_0(z)} &= j e^{j(z-\frac{3\pi}{4}) 
- j(z-\frac{\pi}{4}) } \\
&= j e^{j(-\frac{\pi}{2})} \\
&=1,
\end{split}
\end{equation}
where the functions $A_i$ are defined as in 
\eqref{asymptotic_bessel}.
Multiplying the left-hand side of \eqref{lemma1} 
by \eqref{A-fraction} and computing the limits
with \eqref{limJA}, we find
\begin{equation}
\begin{split}
\lim_{\omega \rightarrow \infty} \frac{J_0(kR)}{J_1(kR)} &= 
\lim_{\omega \rightarrow \infty} \frac{J_0(kR)}{J_1(kR)} 
\frac{jA_1(kR)}{A_0(kR)} \\
&= 
j \lim_{\omega \rightarrow \infty} \frac{J_0(kR)}{A_0(kR)} 
 \lim_{\omega \rightarrow \infty} \frac{1}{\frac{J_1(kR)}{A_1(kR)}} \\
 &= j.
\\
\end{split}
\end{equation}

\section*{Appendix B}
\label{appendixb}
\vspace{-9pt}
We simplify the formula
\begin{equation}
\label{Bvaite}
\frac{\int_L R(l) l g_{\alpha,\beta,l_\text{max}}(l) \text{d}l }
		   {\int_L l g_{\alpha,\beta,l_\text{max}}(l) \text{d}l }.
\end{equation}

Some computations are similar to what has been carried out earlier 
\cite{vesa_eddy-current_2023}.
Let us first compute
\begin{equation}
	\int_L R(l)^i l g_{\alpha,\beta,l_\text{max}}(l) \text{d}l,
\end{equation}
where $i$ is a non-negative integer.
First, we note that $R=l/(2\kappa)$ and 
$g_{\alpha,\beta,l_\text{max}}= 
Cl^{\alpha-4}(l_\text{max}-l)^{\beta-1}$, as defined in
Sections \ref{Section-Preliminaries} and 
\ref{Section-EffectiveRadius}, respectively.
Second, substituting $l=l_\text{max}q$ and 
$\text{d}l=l_\text{max}\text{d}q$, we get
\begin{equation}
\label{Bjohto1}
\begin{split}
& \int_L R(l)^i l g_{\alpha,\beta,l_\text{max}}(l) \text{d}l \\
&=  \frac{1}{(2\kappa)^i} \int_L 
l^{i+1}g_{\alpha,\beta,l_\text{max}}(l) \text{d}l  \\
&=  \frac{1}{(2\kappa)^i} \int_L 
l^{i+1}Cl^{\alpha-4}(l_\text{max}-l)^{\beta-1} \text{d}l  \\
&=  \frac{C}{(2\kappa)^i} \int_L 
l^{i+\alpha-3}(l_\text{max}-l)^{\beta-1} \text{d}l  \\
&=  \frac{C}{(2\kappa)^i} \int_0^1 
(l_\text{max}q)^{i+\alpha-3}
(l_\text{max}-l_\text{max}q)^{\beta-1} l_\text{max}\text{d}q  \\
&=  \frac{C}{(2\kappa)^i} \int_0^1 l_\text{max}^{i+\alpha-3}q^{i+\alpha-3}
l_\text{max}^{\beta-1}(1-q)^{\beta-1} l_\text{max}\text{d}q  \\
&=  \frac{Cl_\text{max}^{i+\alpha+\beta-3}}{(2\kappa)^i} 
\int_0^1 q^{i+\alpha-3} (1-q)^{\beta-1} \text{d}q  \\
&=  \frac{Cl_\text{max}^{i+\alpha+\beta-3}}{(2\kappa)^i} 
\int_0^1 q^{(i+\alpha-2)-1} (1-q)^{\beta-1} \text{d}q.  \\
\end{split}
\end{equation}
The integral is now of the form of the Beta function
\begin{equation}
B(z_1,z_2) = \int_0^1 q^{z_1-1} (1-q)^{z_2-1} \text{d}q,
\end{equation}
which decomposes into Gamma functions as
\begin{equation}
B(z_1,z_2) = \frac{\Gamma(z_1)\Gamma(z_2)}{\Gamma(z_1+z_2)},
\end{equation}
as pointed out in the literature 
\cite{abramowitz_handbook_1972}.
Hence, the integral \eqref{Bjohto1} turns into
\begin{equation}
\label{Bjohto2}
\begin{split}
& \int_L R(l)^i l g_{\alpha,\beta,l_\text{max}}(l) \text{d}l \\
&= \frac{Cl_\text{max}^{i+\alpha+\beta-3}}{(2\kappa)^i}
\frac{\Gamma(i+\alpha-2)\Gamma(\beta)}{\Gamma(i+\alpha+\beta-2)}.
\end{split}
\end{equation}

Let us now compute \eqref{Bvaite}.
Using the identity
\begin{equation}
\Gamma(z+1) = z\Gamma(z),
\end{equation}
appearing in the textbooks \cite{abramowitz_handbook_1972}, 
and applying \eqref{Bjohto2} for $i=0$ and $i=1$, we get
\begin{equation}
\label{Bjohto3}
\begin{split}
&\frac{\int_L R(l) l g_{\alpha,\beta,l_\text{max}}(l) \text{d}l }
		   {\int_L l g_{\alpha,\beta,l_\text{max}}(l) \text{d}l } \\
&=\frac{Cl_\text{max}^{1+\alpha+\beta-3}}{2\kappa}
\frac{\Gamma(1+\alpha-2)\Gamma(\beta)}{\Gamma(1+\alpha+\beta-2)} \\
& \ \ \ \ \ \ \ \    \cdot\frac{1}{Cl_\text{max}^{\alpha+\beta-3}}
\frac{\Gamma(\alpha+\beta-2)}{\Gamma(\alpha-2)\Gamma(\beta)} \\
&=\frac{l_\text{max}}{2\kappa}
\frac{\Gamma(1+\alpha-2)\Gamma(\beta)}{\Gamma(1+\alpha+\beta-2)} \\
& \ \ \ \ \ \ \ \    \cdot
\frac{\Gamma(\alpha+\beta-2)}{\Gamma(\alpha-2)\Gamma(\beta)} \\
&=\frac{l_\text{max}}{2\kappa}
\frac{\alpha-2}{\alpha+\beta-2}
\frac{\Gamma(\alpha-2)\Gamma(\beta)}{\Gamma(\alpha+\beta-2)} \\
& \ \ \ \ \ \ \ \   \cdot
\frac{\Gamma(\alpha+\beta-2)}{\Gamma(\alpha-2)\Gamma(\beta)} \\
&=\frac{l_\text{max}}{2\kappa}
\frac{\alpha-2}{\alpha+\beta-2}.
\end{split}
\end{equation}

\section*{Appendix C}
We show that the conditions \eqref{conditions4} 
and \eqref{conditions5} 
are equivalent,  provided that the used distribution is the 
beta distribution \eqref{betadistr}. 
The associated shape parameters $\alpha$ and
$\beta$ are given by 
\eqref{alpha} and \eqref{beta}.

Let us first show the implication
\begin{equation}
\label{appcimpl1}
\begin{cases}
\alpha > 2,\\
\beta > 0, \\
l_\text{max} > 0,
\end{cases} \Rightarrow \begin{cases}
\sigma_G^2 > 0,\\
\mu_G^2 > 2\sigma_G^2, \\
l_\text{max} > \frac{\mu_G^2+\sigma_G^2}
{\mu_G-\frac{2\sigma_G^2}{\mu_G}}.
\end{cases}
\end{equation}
First of all, let us show that from the 
left conditions of \eqref{appcimpl1}
it follows that $\sigma_G^2>0$. 
The variance of the particle lengths in the particle volume distribution
$G_{\alpha,\beta,l_\text{max}}(l)=
Dl^{\alpha-1}(l_\text{max}-l)^{\beta-1}$ is given by
\begin{equation}
\sigma_G^2 = D\int_0^{l_\text{max}}(l-\mu_G)^2 l^{\alpha-1} 
(l_\text{max}-l)^{\beta-1} \text{d}l.
\end{equation}
It must be that $\sigma_G^2>0$, since the normalization coefficient $D>0$, 
and the integrand is positive 
almost everywhere in $[0,l_\text{max}]$. 
(Similarly, it must be that $\mu_G>0$.)

Let us next show that the condition $\mu_G^2 > 2\sigma_G^2$ 
follows from the left side of \eqref{appcimpl1}.
Using the formula for $\alpha$, \eqref{alpha}, we have
\begin{equation}
\label{appcder1}
\alpha = \frac{(l_\text{max}\mu_\text{G}-\mu_\text{G}^2-\sigma_\text{G}^2)
\mu_\text{G} }{l_\text{max}\sigma_\text{G}^2 } > 2
\end{equation} \vspace{-15pt}
\begin{alignat*}{2}
&\Rightarrow& (l_\text{max}\mu_\text{G}-\mu_\text{G}^2-\sigma_\text{G}^2)
\mu_\text{G} 
&> 2l_\text{max}\sigma_\text{G}^2 \\ 
&\Rightarrow& l_\text{max}\mu_\text{G}^2-\mu_\text{G}^3-\sigma_\text{G}^2
\mu_\text{G} 
&> 2l_\text{max}\sigma_\text{G}^2 \\ 
&\Rightarrow& l_\text{max}\mu_\text{G}^2-2l_\text{max}\sigma_\text{G}^2  
&> \mu_\text{G}^3 + \sigma_\text{G}^2\mu_\text{G} \\ 
&\Rightarrow& l_\text{max}(\mu_\text{G}^2-2\sigma_\text{G}^2)  
&> \mu_\text{G}^3 + \sigma_\text{G}^2\mu_\text{G}. \\ 
\end{alignat*}
Since $l_\text{max},\mu_G,\sigma_G^2>0$, it must be that 
$\mu_\text{G}^2-2\sigma_\text{G}^2>0$,
from which $\mu_G^2 > 2\sigma_G^2$ follows.

Finally, we show  that the condition for $l_\text{max}$ follows from the 
left side of \eqref{appcimpl1}.
From the derivation after \eqref{appcder1}, we may continue as
\begin{alignat*}{2}
&& l_\text{max}(\mu_\text{G}^2-2\sigma_\text{G}^2)  
&> \mu_\text{G}^3 + \sigma_\text{G}^2\mu_\text{G} \\ 
&\Rightarrow& l_\text{max}  
&> \frac{\mu_\text{G}^3 + \sigma_\text{G}^2\mu_\text{G}}
{\mu_\text{G}^2-2\sigma_\text{G}^2} \\ 
&\Rightarrow& l_\text{max}  
&> \frac{\mu_\text{G}^2 + \sigma_\text{G}^2}
{\mu_\text{G}-\frac{2\sigma_\text{G}^2}{\mu_\text{G}}}. \\
\end{alignat*}

Let us next show the implication
\begin{equation}
\label{appcimpl2}
\begin{cases}
\alpha > 2,\\
\beta > 0, \\
l_\text{max} > 0,
\end{cases} \Leftarrow \begin{cases}
\sigma_G^2 > 0,\\
\mu_G^2 > 2\sigma_G^2, \\
l_\text{max} > \frac{\mu_G^2+\sigma_G^2}{\mu_G-\frac{2\sigma_G^2}{\mu_G}}.
\end{cases}
\end{equation}

Let us show that $l_\text{max}>0$ follows from the 
right-hand side conditions of \eqref{appcimpl2}.
From the right-hand side conditions, it follows that 
$\mu_G^2 > 2\sigma_G^2>0$, 
so both $\mu_G$ and $\sigma_G^2$ are positive. 
For $l_\text{max}$, we find that
\begin{equation}
\label{appcder2}
\begin{split}
l_\text{max} &> \frac{\mu_G^2+\sigma_G^2}
{\mu_G-\frac{2\sigma_G^2}{\mu_G}} \\ 
&> \frac{\mu_G^2}{\mu_G-\frac{2\sigma_G^2}{\mu_G}} \\ 
&=\frac{\mu_G^3}{\mu_G^2-2\sigma_G^2} \\
&>\frac{\mu_G^3}{\mu_G^2} =\mu_G>0.
\end{split}
\end{equation}
Now we know that all the parameters, $\sigma_G$, $\mu_G$, and 
$l_\text{max}$ are positive.

Let us next show that $\alpha > 2$ follows from the 
right-hand side conditions of \eqref{appcimpl2}.
We find that
\begin{alignat*}{2}
&& l_\text{max}  
> \frac{\mu_\text{G}^2 + \sigma_\text{G}^2}{\mu_\text{G}-
\frac{2\sigma_\text{G}^2}{\mu_\text{G}}}
&=\frac{\mu_\text{G}^3 + \sigma_\text{G}^2\mu_\text{G}}
{\mu_\text{G}^2-2\sigma_\text{G}^2} \\
&\Rightarrow& l_\text{max}(\mu_\text{G}^2-2\sigma_\text{G}^2)  
&> \mu_\text{G}^3 + \sigma_\text{G}^2\mu_\text{G} \\ 
&\Rightarrow& l_\text{max}\mu_\text{G}^2-2l_\text{max}\sigma_\text{G}^2  
&> \mu_\text{G}^3 + \sigma_\text{G}^2\mu_\text{G} \\ 
&\Rightarrow& l_\text{max}\mu_\text{G}^2-\mu_\text{G}^3-\sigma_\text{G}^2
\mu_\text{G} 
&> 2l_\text{max}\sigma_\text{G}^2 \\ 
&\Rightarrow& (l_\text{max}\mu_\text{G}-\mu_\text{G}^2-\sigma_\text{G}^2)
\mu_\text{G} 
&> 2l_\text{max}\sigma_\text{G}^2, \\ 
\end{alignat*}
from which we conclude that
\begin{equation}
\label{appcder3}
\alpha = \frac{(l_\text{max}\mu_\text{G}-\mu_\text{G}^2-
\sigma_\text{G}^2)\mu_\text{G} }
		{l_\text{max}\sigma_\text{G}^2 } > 2.
\end{equation} 

Finally, we show that $\beta > 0$ follows from the right-hand side 
conditions of \eqref{appcimpl2}.
From \eqref{appcder2}, we find that 
\begin{equation}
l_\text{max}-\mu_G>0.
\end{equation}
From \eqref{appcder3}, we find that
\begin{equation}
\frac{(l_\text{max}\mu_\text{G}-\mu_\text{G}^2-\sigma_\text{G}^2) }
		{l_\text{max}\sigma_\text{G}^2 } > \frac{2}{\mu_\text{G}}>0.
\end{equation} 
Hence, to complete the proof, we have
\begin{equation}
\beta = \frac{(l_\text{max}\mu_\text{G}-\mu_\text{G}^2-\sigma_\text{G}^2)
(l_\text{max}-\mu_\text{G}) }
		{l_\text{max}\sigma_\text{G}^2 } > 0.
\end{equation}

\printbibliography

\end{document}